\documentclass[conference]{IEEEtran}
\usepackage{cite}
\usepackage{amsmath,amssymb,amsfonts}
\usepackage{algorithmic}
\usepackage{graphicx}
\usepackage{textcomp}
\usepackage{xcolor}
\usepackage{hyperref}

\definecolor{cardinal}{rgb}{0.77, 0.12, 0.23}
\definecolor{officegreen}{rgb}{0.0, 0.5, 0.0}
	\definecolor{lightbrown}{rgb}{0.71, 0.4, 0.11}

\usepackage{multirow}
\usepackage{adjustbox}
\usepackage{makecell}
\usepackage{subcaption}

\def\BibTeX{{\rm B\kern-.05em{\sc i\kern-.025em b}\kern-.08em
    T\kern-.1667em\lower.7ex\hbox{E}\kern-.125emX}}
\begin{document}

\title{Vax-Culture: A Dataset for Studying Vaccine Discourse on Twitter\\
}

\author{\IEEEauthorblockN{Mohammad Reza Zarei \IEEEauthorrefmark{1}, Michael Christensen \IEEEauthorrefmark{2}, Sarah Everts \IEEEauthorrefmark{3} and Majid Komeili \IEEEauthorrefmark{4}}
\IEEEauthorblockA{\IEEEauthorrefmark{1} \IEEEauthorrefmark{4} School of Computer Science, \IEEEauthorrefmark{2} Department of Law and Legal Studies, \IEEEauthorrefmark{3} School of Journalism and Communication} 
\textit{Carleton University}\\
Ottawa, Canada \\
Email: \IEEEauthorrefmark{1}Mohammadrezazarei@cmail.carleton.ca,
\IEEEauthorrefmark{2}Michael.Christensen@carleton.ca,
\IEEEauthorrefmark{3}Sarah.Everts@carleton.ca,\\
\IEEEauthorrefmark{4}Majid.Komeili@carleton.ca}

\maketitle

\begin{abstract}
Vaccine hesitancy continues to be a main challenge for public health officials during the COVID-19 pandemic. As this hesitancy undermines vaccine campaigns, many researchers have sought to identify its root causes, finding that the increasing volume of anti-vaccine misinformation on social media platforms is a key element of this problem. We explored Twitter as a source of misleading content with the goal of extracting overlapping cultural and political beliefs that motivate the spread of vaccine misinformation. 
To do this, we have collected a data set of vaccine-related Tweets and annotated them with the help of a team of annotators with a background in communications and journalism. 
Ultimately we hope this can lead to effective and targeted public health communication strategies for reaching individuals with anti-vaccine beliefs. Moreover, this information helps with developing Machine Learning models to automatically detect vaccine misinformation posts and combat their negative impacts. In this paper, we present Vax-Culture, a novel Twitter COVID-19 dataset consisting of 6373 vaccine-related tweets accompanied by an extensive set of human-provided annotations including vaccine-hesitancy stance, indication of any misinformation in tweets, the entities criticized and supported in each tweet and the communicated message of each tweet. Moreover, we define five baseline tasks including four classification and one sequence generation tasks, and report the results of a set of recent transformer-based models for them.
The dataset and code are publicly available at \url{https://github.com/mrzarei5/Vax-Culture}.

\end{abstract}


\section{Introduction}

When the novel coronavirus known as COVID-19 emerged in 2019, it spread rapidly enough to be classified as a global pandemic by the World Health Organization in early 2020. Almost as rapidly, medical researchers began developing vaccines, culminating in  multiple effective options authorized by national drug safety bodies around the world. To decrease morbidity and to catalyze population-wide immunity, public health agencies in many countries began to launch vaccine campaigns. However, despite the wide availability in some countries, a subset of many populations avoided vaccination, a phenomenon known as vaccine hesitancy \cite{nyawa2022covid}.

Vaccine hesitancy undermines important public health campaigns.  Over the past decade,  social scientists have found that an increasing volume of anti-vaccine misinformation and disinformation spread online is a key element of this problem \cite{featherstone2020feeling,jamison2020adapting,smith2019mapping}. Specifically, social media platforms have been widely used to spread anti-vaccine misinformation. These platforms are among the main tools connecting people (particularly in a pandemic) and they have significant influence on decision-making. Yet, as social scientists have pointed out, misinformation and conspiracy theories are not spread by people who are simply uninformed. For many, misinformation can tap into deep cultural narratives or the stories that people use to make sense of their lives and social contexts. Misinformation can therefore feel true because it is interesting and entertaining \cite{hochschild2018strangers,polletta2019deep}, or even products of what believers see as critical thinking and research \cite{boyd2018you}. From this perspective, if misinformation feels true because it speaks to social anxieties or seems more consistent with peoples’ political identities, combating misinformation may not simply be a problem of education and fact-checking. Therefore, successful communication strategies for combating vaccine hesitancy require a sophisticated understanding of the underlying factors motivating misinformation spread.

In one of the most recent broad manifest of vaccine hesitancy, social media platforms were widely used to propagate misinformation and negative tendencies against COVID-19 vaccines. Hence, these sources can be used to explore COVID-19 vaccine misinformation posts and search for clues about how best to reach anti-vaccine or vaccine hesitant members of the public and understand the cultural/media context in which hesitancy has grown. Moreover, the availability of high-quality information in this area can aid in developing Machine Learning models for automatically identifying vaccine misinformation posts and their attributes that can lead to combating their negative impacts. However, reaching these objectives depends on collecting relevant data.

Numerous Covid-19 vaccine datasets have been recently collected from social media and made publicly available. These datasets are mainly gathered on the Twitter platform due to the simplicity of working with Twitter API and the wide range of functionalities this tool provides for pulling Tweets. Moreover, Twitter is one of the most important sources of misleading content in the media ecosystem \cite{hindman2018disinformation,phillips2018oxygen} making it the center of attention in this area. 

Although the published datasets have expedited the access of researchers to vaccine-related data, they are usually collected and published without any quality control. Furthermore, annotations regarding vaccine-related attributes such as vaccine hesitancy and any misleading information in the text are missing from these datasets, making them unsuitable for training models on supervised tasks. Moreover, this missing information adds an additional process for performing studies that require annotation. Finally, since only tweet IDs can be published due to Twitter policies and these Twitter datasets are not published with additional information regarding the meaning or communicated message in each tweet, if the majority of the tweets get deleted over time, the dataset will become useless. 

In this paper, we present a Twitter COVID-19 vaccine misinformation dataset (Vax-Culture) pulled from Twitter using the Twitter API with vaccine-related keywords, then annotated by a team of trained annotators with a background in journalism and communication. Each tweet is accompanied by a set of complete attributes including vaccine hesitancy stance (whether a Tweet is pro- or anti-vaccine), whether any information is misleading or inaccurate in the tweet, and the entities that are supported or criticized in the tweet. Since a small proportion of tweets in Twitter datasets might get deleted over time, we also provide the intended communicated meaning of each tweet, mitigating the problem of deleted tweets. Moreover, we define five baseline tasks including two multi-class classification tasks, two multi-label classification tasks, and one text generation task on our dataset, and report the performance of a set of state-of-the-art transformer-based baseline models on these tasks. This evaluation can give insights to researchers about developing models using our dataset.

\section{Related Work}

Since the advent of COVID-19 virus, many Twitter datasets related to this disease have been collected and published. Lamsal \cite{lamsal2021design} presented a COVID-19 dataset consisting of 310 million Covid-19 English language tweets collected with covid-related keywords starting from March 20, 2020. The sentiment of each tweet was also presented which was computed using TextBlob. A multilingual COVID-19 dataset was introduced in \cite{chen2020tracking}, comprised of 123 million tweets, with over 60\% of the tweets in English by the time of publishing its paper.  A set of basic analyses was also presented in the paper that showed a correlation between coronavirus-related events and Twitter activity. The first tweet of the presented dataset backs to January 21, 2020 and its collection is still in progress. Another large-scale multilingual COVID-19 dataset was presented in \cite{lopez2021augmented} containing a total of 2,996,610,622 tweets as of November 30, 2022 that were collected using Twitter’s trending topics and selected keywords. This dataset is augmented with Twitter Named Entity Recognition and Sentiment Analysis algorithms. In \cite{gupta2020covid}, a dataset consisting of 252,600,524 tweets was presented, collected from January 28, 2020 to June 1, 2022 and published with the emotion and sentiment of each tweet. This additional information was extracted using pre-trained Machine Learning-based emotion recognition algorithms. Topic Modeling was also performed to extract the topic of each tweet.

All of the above datasets, however, either do not have any annotation for the tweets or are accompanied by information that is obtainable without any manual annotation such as topic, sentiment and emotion. Such annotations are not directly related to the COVID-19 vaccine subject. Therefore, they could not be directly used for tasks such as vaccine hesitancy detection that requires customized annotations. Moreover,  In \cite{hussain2021artificial}, a dataset with respect to the context of vaccination for the COVID-19 pandemic was presented, consisting of 40,268 and 98,385 tweets from the United Kingdom and the United States, respectively. Although the sentiment of a tweet may not necessarily be an indicator of vaccine hesitancy, this information is directly used to specify whether the communicated message in each tweet is vaccine-hesitant or not. COVID-19 vaccine hesitancy detection with human-provided annotations has also been considered in recent research. However, most of these papers have not made their annotated datasets publicly available \cite{nyawa2022covid,cotfas2021covid,cotfas2021longest}. 

In \cite{hayawi2022anti}, a Twitter dataset for vaccine misinformation detection was introduced. This dataset comprised a total of 15,073 tweets, with 5,751 labeled as misinformation. The tweets were annotated using verified sources and then the labels were validated by public health experts. Three different vaccine misinformation detection models including XGBoost, LSRTM, and BERT were trained and evaluated on the collected dataset. Although the spread of misinformation can be a sign of vaccine hesitancy, the presence or absence of misinformation alone is not adequate for developing models for vaccine hesitancy detection. Our collected dataset contains an extensive set of annotations related to vaccine misinformation and vaccine hesitancy that enables performing a thorough analysis regarding the spread of vaccine misinformation and understanding the context in which hesitancy has grown. Moreover, this dataset opens the door for developing Machine Learning-based models on several natural language processing tasks related to vaccine misinformation including vaccine hesitancy detection, vaccine misinformation detection, criticized and supported entities detection as well as tweet communicated meaning generation. In the paper, we focus on the latter, as the former goes beyond the machine learning and natural language processing area and we leave it for future work.

\section{Dataset Collection and Annotation}
In this section, we describe the data collection process using Twitter API and the tweet annotation process performed by a trained team of annotators with a background in Journalism and Communication.

\textbf{Tweet data collection:} 
We collected vaccine-related tweets containing specific keywords using Twitter API for seven months from September 2021 to March 2022 with a rate of ~5000 tweets every day. Each tweet contained at least one of the following keywords: vaccine, vaccines, vaccination, vacc, vax, vaxx and vaxxed. We further processed the collected tweets to remove duplicates and non-English language tweets, resulting in a collection of ~500k tweets.

\textbf{Annotation Process:} 
A web dashboard was created to annotate tweets and manage assignment of tweets to annotators. The dashboard allows annotators to assess an individual tweet as it appears on the Twitter platform and annotate it according to the following criteria: (1) whether it is anti-vaccine, neutral, or pro-vaccine, (2) whether any information in the tweet seems misleading or inaccurate (3) a list of who or what is the subject of criticism in the tweet, (4) a list of who or what is supported or promoted in the tweet (5) A free text response explaining the intended meaning of the tweet.

\begin{table*}
\caption{List of the questions and possible answer choices used to annotate each tweet. For questions 3 and 4, in addition to the predefined set of answer choices, annotators were provided with the possibility of entering additional answers in a free-text response form.}
\label{tab_1}
\centering

\begin{tabular}{l|l}
\hline
Question & Answer Options \\
\hline
1. What is the message communicated in this Tweet? & 1. Anti-vaccine 2. Pro-vaccine 3. Unsure about the vaccine \\
\hline
2. Does any information in the Tweet seem misleading or inaccurate? & 1. Yes 2. No\\
\hline
\multirow{5}{*}{3. Who or what is the subject of criticism in the tweet? [choose all that apply]} & 1. Vaccine mandates 2. Anti-vaxxers 3. The safety of vaccines \\

 & 4. Vaccine effectiveness 5. Public health policy 6. Politicians\\
 & 7. Government 8. Public health officials 9. Pharmaceutical companies\\
 & 10. Democrats or Liberals 11. Conservative media 12. Mainstream \\
 & media (Additional answers in the form of free text can be provided)\\
\hline
\multirow{5}{*}{4. Who or what does the tweet support or promote? [choose all that apply]} & 1. Vaccines 2. Freedom of choice 3. Public health interventions \\ 
& 4. A more relaxed approach 5. Science 6. Natural health \\
& 7. Global response 8. Waiting for more information 9. Alternative \\
& remedies 10. Small business 11. Religious beliefs (Additional answers \\
& in the form of free text can be provided)\\

\hline

5. Please explain the meaning this tweet intends to communicate (either &\\
implicitly or explicitly). Please also include any important keywords related & Free text answer should be provided by annotator \\to this meaning. &\\

\end{tabular}
\end{table*}

This evaluation is performed by answering five questions, each reflecting one of the mentioned criteria. For question 2, annotators were not asked to verify the factual accuracy of any given tweet, but only to evaluate the presentation of the information. The complete list of questions and possible answer choices are shown in Table \ref{tab_1}. Since there may be an entity criticized or promoted in the corresponding tweet which is not among the provided answer options in questions 3 and 4, we let the annotators add additional entities by considering a free-text box for questions 3 and 4 to enter other answers. The intended meaning of each tweet is requested in question 5 to first complement the text of each tweet for analysis tasks, and second to keep some records and information about the content of each tweet, in case the tweet was deleted.

The process of identifying misinformation is highly subjective, and therefore requires a baseline literacy of the topic under consideration. Our early efforts to test the annotation system using the average score of three annotators on Amazon's MTurk platform \footnote{https://www.mturk.com/} returned very poor results. As such we adopted a quality control protocol focused on hiring annotators with existing English-language media literacy, in this case, graduate students and fourth-year advanced undergraduate students from Carleton University's School of Journalism and Communication. We then hosted three training sessions with annotators designed to ensure interpretive alignment for each of our questions on the topic of vaccines, as well as our expectations for the free-text annotations (including minimum word count, level of generality, and a focus on subtextual interpretation). We also spot-checked the work of each annotator by monitoring word counts, time per annotation, and answer frequency for each question. The results made clear that one trained annotator produced far richer and more accurate data than the crowd-sourced annotators were able to produce. 

Each annotator is assigned a disjoint subset of random tweets from the tweets pool for annotation. Since the content of all tweets pulled by keywords may not be relevant, we allow the annotators to skip any of the tweets if they found the content is not informative. Overall, 6,373 tweets were annotated by the team of our annotators that forms our dataset. The main statistics of the annotated dataset are reported in Table \ref{tab_2}. According to the statistics, 43.8\% and 40.5\% of the tweets are anti-vaccine and pro-vaccine, respectively. This shows a good balance between positive and negative tendencies over the annotated tweets with respect to communicated messages. Also, 15.8\% of the tweets communicate uncertainty about vaccine. Moreover, 38.9\% of the tweets contain misleading or inaccurate information. Finally, 77.3\% of tweets criticise at least one of the possible answer options such as vaccine mandate, anti-vaxxers, safety of vaccines, etc.. This number is 87\% for supporting entities such as Vaccines, freedom of choice, public health interventions, etc.. We present example pairs of tweets and their corresponding communicated meaning of the tweets that are collected by annotators in Table \ref{tab_t_meaning}. 

\begin{table}
\caption{Statistics of the dataset}
\label{tab_2}
\centering

\begin{tabular}{l|l}
\hline
Number of tweets & 6,373\\
Percentage of anti-vaccine tweets & 43.8\% \\
Percentage of pro-vaccine tweets & 40.5\% \\
Percentage of tweets with uncertainty about vaccine & 15.8\%\\
Percentage of tweets with misleading/inaccurate information & 38.9\%\\
Percentage of tweets criticizing at least one entity & 77.3\%\\
Percentage of tweets supporting at least one entity & 87\%\\

\end{tabular}
\end{table}



\begin{table*}
\caption{Examples of tweets and the meaning they intend to communicate, as provided by the annotators.}
\label{tab_t_meaning}
\centering

\begin{tabular}{cc}
\hline
Tweet & Meaning \\
\hline
\makecell{@USER All these conspiracy theorists are more likely to get seriously\\ ill, plus there’s more vaccine for the thinking population. We call that\\ win/win.} & \makecell{The tweet is promoting the fact that anti-vaxxers will get sick from not\\ getting vaccinated. The tweet is in support of this, and specifies that\\ there will be more vaccines for the thinking population.}\\
\hline
\makecell{When a totalitarian government and big pharma have a lovechild you\\ get mandatory vaccination.} & \makecell{This tweet is saying that governments and big pharma are working together\\ and that by creating vaccine mandates they are both benefitting,\\ which is not true.}\\
\hline

\makecell{So people who have never had COVID are getting the vaccine and\\ now they are coming down with COVID. Anyone care to explain that?} &
\makecell{The tweet is questioning the safety of vaccines, as he is skeptical that\\ people who are getting the vaccine, are now contracting COVID-19\\ as well.}\\
\hline
\makecell{So many questions for @USER. If the vaccine is safe and effective,\\ why won't you, doctors, manufacturers take responsibility if \\something goes wrong? HTTPURL} & 
\makecell{This tweet is referring to the idea that pharmaceutical manufacturers\\ are not responsible for any vaccine-related deaths and includes a video\\ claiming that 40\% of the people who are dying from COVID are\\ completely vaccinated. The tweeter is pointing to the idea that if\\ officials truly believed vaccines were safe and effective then they \\would take accountability when things go wrong.}\\

\end{tabular}
\end{table*}

\section{Baseline Tasks}
\label{sec_base}

In this section, we introduce a set of baseline tasks.

\textbf{Communicated message prediction:}
This task consists of predicting the communicated message of each tweet and is defined as a three-way classification task: anti-vaccine, pro-vaccine or unsure about vaccine.

\textbf{Misleading or inaccurate information detection:}
This task is defined as a binary classification and consists of predicting whether any information in the tweet seems misleading or inaccurate.

\textbf{Subjects of criticism prediction:}
This task is defined as a multilabel classification problem and consists of predicting all entities that have been criticized. This task is a 12-class multilabel classification problem. The available classes are the same as the answer options for question 3 as shown in Table \ref{tab_1}.

\textbf{Subjects of support/promote prediction:}
Similar to the previous tasks, this one is also a multilabel classification problem. It consists of predicting the entities that each tweet supports or promotes. This task consists of 11 classes which are the same as the answer options for question 4 as shown in Table \ref{tab_1}. 

\textbf{Tweet meaning generation:}
This task is defined as a text generation task to explain the meaning each tweet intents to communicate either implicitly or explicitly as provided by the annotators in response to question 5. The input is the text of the corresponding tweet. Since the text of tweet may not be sufficient for this task, we consider two additional versions of this task with extended inputs. In the first extended version, we append the text of the replied tweet in case the tweet is a reply. This includes 31.1\% of all the tweets. The new tweet is preceded with the token ``$<rep>$'' in the input to separate it from the text of the original tweet. We also append the title and description of the links that were referred in tweet and parsed by Twitter API. This was applicable to 13.3\% of tweets. The titles are preceded with ``$<url\_t>$'' and the descriptions are preceded with ``$<url\_d>$''. In the second extended version, we expand the input by adding the information collected by the annotators from questions 1 to 4. To this end, we transform the answers for each tweet to sentences and append them to the original tweet to form an extended input text. This information is separated from the tweet text with the token $<sep>$. The following outlines the transformation process for each question:

\begin{itemize}
   \item[Q1.] The template sentence ``The message communicated in this tweet is . . .'' is completed with the answer.
   \item[Q2.] If the answer to this question is positive, the sentence ``This tweet contains misleading or inaccurate information.'' is appended to input
   \item[Q3.] For any option selected by the annotators, we add a sentence by completing the template ``This tweet criticises . . .''.
   \item[Q4.] Similar to the previous question, we append a sentence for any selected option by completing the template ``This tweet supports or promotes . . .''.
\end{itemize}

\section{Baseline Tasks Evaluation}

\subsection{Experimental Setup}

\textbf{Classification Models:}
We used five different pre-trained language models for baseline classification tasks (communicated Message prediction, misleading/inaccurate information detection, subjects of criticism prediction and subject of support/promote prediction) including BERT-base \cite{devlin2018bert}, BERT-large \cite{devlin2018bert}, RoBERTa-base \cite{liu2019roberta}, RoBERTa-large \cite{liu2019roberta} and BERTweet-covid19 \cite{nguyen2020bertweet}. BERT-base and BERT-large are pre-trained on BookCorpus \cite{zhu2015aligning} and English Wikipedia. RoBERTa-base and RoBERTa-large are pre-trained on over 160GB of uncompressed text from BookCorpus, English Wikipedia, CC-News containing 63 million English news articles collected from the CommonCrawl News dataset \cite{nagel2016cc}, OpenWebText \cite{gokaslan2019openwebtext} and Stories \cite{trinh2018simple}. BERTweet-covid19 has the same architecture as BERT-base but is specifically pre-trained on English tweets. This model uses 850M English tweets plus 23M COVID-19 English tweets for pre-training.

\textbf{Text Generation Models:}
To perform baseline experiments for tweet meaning generation task, we used Bart-large \footnote{https://huggingface.co/facebook/bart-large} \cite{lewis2019bart} and T5-large \cite{raffel2020exploring} \footnote{https://huggingface.co/t5-large} which are two large-scale language models suitable for text generation tasks.

\textbf{Data Split:}
We use the same train, validation, and test sets in all experiments. To create these sets, we perform a data split with a stratified sampling over the field "communicated massage" to ensure the same distribution in terms of vaccine hesitancy along all train, validation and test sets. We use 20\% of the dataset as test set and keep 25\% of the remaining tweets to perform validation and select the best model in each experiment. The remainder is used for training.

\textbf{Experimental Settings:}
For all experiments, we substitute mentions with the token ``@USER'' and URL links with ``HTTPURL'' in all tweets. Moreover, we use the emoji package of Python to replace each emoji with a text representation. We also use NLTK TweetTokenizer \cite{bird2009natural} to remove redundant characters. 

For all tasks, the tokens ``HTTPURL'' and ``@USER'' are added to the vocabulary of each model before being fine-tuned on the dataset. Additional to these two tokens, we add the tokens $<rep>$, $<url\_t>$, and $<url\_d>$ for the tweet meaning generation task with the first version of extended input. For the second version, we add the token $<sep>$ to the vocabulary of the corresponding models.

\begin{table}
\caption{Evaluation of the baseline models on the communicated message prediction task}
\label{tab_t1}
\centering
\begin{tabular}{l|cccc}
\hline
Model & Accuracy & Precision & Recall & F1 \\

\hline
BERTweet-cov19 & 63.2 & 57.5 & 57.6 & 57.3\\
BERT-base & 60.1 & 53.8 & 53.8 & 53.7\\

BERT-large & 60.8 & 54.3 & 54 & 54\\

RoBERTa-base & 60.2 & 55.6 & 55.6 & 55.2 \\

RoBERTa-large & 69.6 & 61.9 & 61.8 & 61.8 \\

\end{tabular}

\end{table}

\begin{table}
\caption{Results of the baseline models on the misleading/inaccurate information detection task.}
\label{tab_t2}
\centering
\begin{tabular}{l|cccc}
\hline
Model & Accuracy & Precision & Recall & F1 \\

\hline
BERTweet-cov19 & 73.4 & 65.2 & 71.8 & 68.3 \\
BERT-base & 71 & 61.7 & 72.4 & 66.6 \\

BERT-large & 71.5 & 61.8 & 75.7 & 68 \\

RoBERTa-base & 72.9 & 66.3 & 65.7 & 66\\

RoBERTa-large & 75.6 & 68 & 73.7 & 70.7\\

\end{tabular}

\end{table}

\textbf{Training Settings:}
Adam optimizer with a learning rate of $1e^{-5}$ and a weight decay of 0.01 is used to fine-tune all classification models. The same optimizer with a learning rate of $3e^{-4}$ and $1e^{-5}$ and no weight decay is used to fine-tune T5-large and Bart-large, respectively. All the models are fine-tuned for 40 epochs with a batch size of 16. After each epoch, the model is evaluated on the validation set and the model with the best performance is evaluated on the test set.

\textbf{Evaluation Metrics:}
We report macro-averaged precision, recall, and F1 score for the classification tasks with the exception of the binary task of misleading/inaccurate information detection where we report precision, recall, and F1 over the class with label "Yes". Tweet meaning generation task is evaluated by three ROUGE metrics including ROUGE-1, ROUGE-2, and ROUGE-L. The length of the generated text is also reported in this task.

\subsection{Baseline Results}

\textbf{Communicated Message Prediction:} The results of this task are shown in Table \ref{tab_t1}. In all cases, the difference between precision and recall was negligible leading to obtaining a value for F1 score close to both precision and recall. The best performance belongs to RoBERTa-large with respect to all metrics. The performance of this model was 69.6 and 61.8 in terms of accuracy and F1 score, respectively. This superiority is due to using a more effective pre-training procedure and a larger set of data for pre-training, compared to Bert models, and a larger network compared to RoBERTa-base. BERTweet-covid19 which is a baseline specifically pre-trained on English tweets ranked second among top-performing models. This model achieved an accuracy of 63.2 and F1 score of 57.3.

\textbf{Misleading/Inaccurate Information Detection:} We present the results of this task in Table \ref{tab_t2}. Similar to the communicated message prediction task, RoBERTa-large outperformed other baselines in terms of all metrics with the exception of best recall score that was attained by BERT-large with 2\% margin.
BERTweet-covid19 is the runner-up with respect to accuracy and F1 score. Although BERTweet-covid19 placed third among best-performing models on both precision and recall, a more balanced performance with respect to these two metrics led to achieving the second highest F1 score with just 0.3\% difference with BERT-large as the third top-performer. BERTweet-covid19 is the runner-up also in terms of accuracy.

\begin{table}
\caption{Evaluating baselines on the subjects of criticism prediction task.}
\label{tab_t3}
\centering
\begin{tabular}{l|cccc}
\hline
Model & Accuracy & Precision & Recall & F1 \\

\hline
BERTweet-cov19 & 88.1 & 47.9 & 28.9 & 33.9 \\
BERT-base & 88.2 & 54.5 & 25.3 & 31.7\\

BERT-large & 88.7 & 57 & 32 & 38\\

RoBERTa-base & 88.2 & 48.8 & 31.9 & 37\\

RoBERTa-large & 89.2 & 51.6 & 41.2 & 44.7\\

\end{tabular}

\end{table}

\begin{table*}
\caption{Per-label F1 score and the ratio of positive class for each answer choice in the Subjects of Criticism task.}
\label{tab_t4}
\centering

\begin{tabular}{l|ccccc|c}
\hline
Label & BERTweet & Bert-base & Bert-large & RoBERTa-base & RoBERTa large & \%Pos Class\\           
\hline
Politicians & 44.8 & 36.5 & 37.7 & 47.9 & 52.8& 10.6\\
Pharmaceutical companies & 39.7 & 38.2 & 53.5 & 44.9 & 52.4 & 5.8\\
Public health officials & 11.7 & 6.1 & 20.5 & 23 & 29.3 & 6.4\\
Anti-vaxxers & 40.8 & 34.8 & 38 & 35.1 & 52.1 & 16.9 \\
Vaccine mandates & 54.7 & 53.6 & 58.2 & 55.1 & 64 & 33.6 \\
Vaccine safety & 47.7 & 49.5 & 49.5 & 51.9 & 62.2 & 18 \\
Conservative media & 8 & 8.3 & 8.3 & 8 & 7.7 & 1.3 \\
Mainstream media & 11.6 & 15.2 & 29.3 & 27.3 & 35.6 & 4 \\
Public health policy & 17.9 & 15.9 & 17.3 & 20.7 & 24.4 & 9.8 \\
Democrats or liberals & 42 & 44.7 & 50.4 & 37.1 & 46.2 & 5.6\\
Government & 44.2 & 34.7 & 43.1 & 45.1 & 50.3 & 13.1\\
Vaccine effectiveness & 43.6 & 42.6 & 50 & 47.6 & 55.6 & 17.6\\

\hline
Average & 33.9 & 31.7 & 38 & 37 & 44.7 & 11.9 \\

\end{tabular}

\end{table*}

\textbf{Subjects of Criticism Prediction:} The results of this task are shown in Table \ref{tab_t3}. RoBERTa-large obtained the highest accuracy, recall and F1. The best performance in terms of precision belongs to BERT-large. This model also holds the second highest accuracy and F1. We observed a considerable gap between the accuracy performance and other metrics. While the baselines were able to achieve almost a high accuracy of around 88\%, they could not perform comparably well on other metrics. To further investigate the results, we present fine-grained F1 scores and the ratio of the positive class for each answer choice in Table \ref{tab_t4}. 

We can observe that the F1 score on some labels including conservative media, mainstream media, public health officials, and public health policy is noticeably low for all baseline models. If we take into consideration the ratio of positive labels, we can see that the ratio of positive labels is below 10\% for these labels which makes them severely imbalanced problems. Although we can see an exception F1 score higher than the average for the label Democrats or liberals that has only 5.6\% instances with positive class, this performance is due to the fact the baselines are not trained from scratch and the pre-trained models perform differently depending on the data used for pre-training. Despite such exceptions in labels with low positive class ratio, we can see that for labels with a more balanced class ratio such as vaccine mandates, performance is significantly higher.

Vaccine mandates was the label with the highest proportion of positive class. This phenomenon was criticised in about one third of the tweets. Next most criticised entities were vaccine safety, vaccine effectiveness and anti-vaxxers with a criticizing ratio of 18\%, 17.6\% and 16.9\%, respectively. Also, conservative media and mainstream media were the rarest among critisized entities with a positive class ratio of 1.3\% and 4\%, respectively.

\textbf{Subjects of Support/Promote Prediction:}
The results of this task are shown in Table \ref{tab_t5}. Similar to the subjects of criticism prediction task, we also report per-label F1 score and the ratio of positive class for each label in Table \ref{tab_t6}. According to Table \ref{tab_t5}, the best performance with respect to all four metrics belongs to RoBERTa-large. This model attained a macro-averaged F1 score of 32.8. Similar to the subjects of criticism prediction task, the macro-average accuracy is around 88 in all cases. This high value is mainly due to the spareness of positive classes where models achieve a high accuracy by simply predicting all samples to be from the majority class. The imbalanced essence of this task should be taken into account to pick the proper model and training procedure. According to Table \ref{tab_t6}, the correlation between low performance in terms of F1 score and the low ratio of the samples with positive class is more tangible as the labels with an F1 score equal to 0 such as religious beliefs and small business hold a positive label ratio less than 5. The highest F1 score for all baselines is achieved for vaccine label that holds the highest ratio of positive class (i.e. 36.2\%) among all labels.

\begin{table}
\caption{Evaluating baselines on the subject of support prediction task.}
\label{tab_t5}
\centering
\begin{tabular}{l|cccc}
\hline
Model & Accuracy & Precision & Recall & F1 \\

\hline
BERTweet-cov19 & 88.3 & 38 & 15.4 & 18.6\\
BERT-base & 88.5 & 38.5 & 15.8 & 20.3\\

BERT-large & 88.7 & 37.4 & 19.8 & 23.6\\

RoBERTa-base & 88.4 & 34.6 & 19.5 & 24\\

RoBERTa-large & 88.8 & 38.6 & 29.5 & 32.8\\

\end{tabular}

\end{table}

\begin{table*}
\caption{Per-label F1 score and the ratio of positive class for each answer choice in the Subjects of Support/Promote task.}
\label{tab_t6}
\centering

\begin{tabular}{l|ccccc|c}
\hline
Label & BERTweet &  Bert-base & Bert-large & RoBERTa-base  & RoBERTa large & \%Pos Label\\          
\hline
Science & 20.1 & 24 & 26 & 28.4 & 33.5 & 17.7\\
Freedom of choice & 36.6 & 37.5 & 43.5 & 45.5 & 52.5 & 31.3\\
Natural health & 0 & 0 & 0 & 12.2 & 21.6 & 2.7 \\
Vaccines & 62.1 & 60.2 & 61.3 & 62.7 & 71.1 & 36.2 \\
Small business & 0 & 0 & 0 & 0 & 0 & 0.4 \\
Alternative remedies & 8.3 & 13.8 & 15.4 & 20.7 & 28.6 & 1.9 \\
A more relaxed approach & 9.7 & 16.4 & 16.7 & 15.5 & 21.8 & 17.3\\
Waiting for more information & 3.6 & 7.1 & 13.1 & 5.6 & 17.3 & 3.1\\
Public health interventions & 28.6 & 32.2 & 35.2 & 31.5 & 44.9 & 19.9 \\
Global response & 35.3 & 31.8 & 48.1 & 41.7 & 50 & 2.9 \\
Religious beliefs & 0 & 0 & 0 & 0 & 20 & 0.7 \\

\hline
Average & 18.6 & 20.3 & 23.6 & 24 & 32.8 & 12.2 \\

\end{tabular}

\end{table*}

\textbf{Tweet Meaning Generation:}
The results of this task are presented in Table \ref{tab_t7}. As explained in the previous section, we consider three versions of the tweet meaning generation task with respect to model input. The simple version that its input is just the text of the tweet is denoted by V0. We also report the results of the models with the first and second versions of the extended input (denoted by V1 and V2, respectively) as explained in Section \ref{sec_base}. 

Although it was anticipated that the models with extended input versions, V1 and V2, would perform better, in our experiments they did not improve V0. The first and the second extended inputs were able to improve the performance of Bart by just 0.7\% and 1.2\% on average, respectively (Average of Rouge-1, Rouge-2, and Rouge-l). Conversely, the average performance of T5 declined by 0.7\% and 0.3\% on V1 and V2, respectively. Through a qualitative analysis of the texts generated by different versions of the models, we observed that none of them was a clear winner in all cases. An example of the input to each version of T5 and the generated meaning is shown in Table \ref{tab_t8}. Although some useful information was inferred and meaningful texts were generated from the tweets, the complex nature of this task led to generating some incomplete meanings in some cases.

\begin{table*}
\caption{The results of tweet meaning generation task using Bart and T5 with different input as explained in Section \ref{sec_base}}
\label{tab_t7}
\centering

\begin{tabular}{l|ccc|ccc|ccc|ccc}
\hline
\multirow{2}{*}{Model} & \multicolumn{3}{c|}{Rouge-1} & \multicolumn{3}{c|}{Rouge-2} & \multicolumn{3}{c|}{Rouge-l} & \multicolumn{3}{c}{Length} 
                     \\

& V0 & V1 & V2 & V0 & V1 & V2 & V0 & V1 & V2 & V0 & V1 & V2\\

T5 & 28.9 & 28 & 28.5 & 8.6 & 8 & 8.5 & 24 & 23.2 & 23.5 & 18.8 & 18.8 & 18.9\\

Bart & 27.4 & 28.3 & 28.6 & 8.5 & 9 & 9.8 & 23.1 & 23.8 & 24.2 & 19.9 & 19.9 & 20\\

\end{tabular}
\end{table*}

\begin{table*}
\caption{An example of input and generated text of T5 model with different versions of input for Tweet Meaning Generation task. The first row corresponds to the model with just the text of the tweet as input. The second and the third rows belong to the model with the first and second versions of extended input, respectively.}
\label{tab_t8}
\centering

\begin{tabular}{cc}
\hline
Input & Output \\
\makecell{@USER Also Neil I love the way you throw about the word\\  vaccines, I know vaccines work.. I am pro vaccine.. My problem\\ is specifically the Covid19 jab.. I know ye like to throw everyone \\in the anti vaxx bucket..} & \makecell{This tweeter says they are pro vaccines and that they are against the COVID-19}\\
\hline
\makecell{@USER Also Neil I love the way you throw about the word\\  vaccines, I know vaccines work.. I am pro vaccine.. My problem\\ is specifically the Covid19 jab.. I know ye like to throw everyone \\in the anti vaxx bucket.. $<rep>$ @USER The truth is vaccines\\ work, I talk to them on a personal level. They are passionate \\about their work. Believe me there’s no conspiracy from science.\\ Pharma makes profit but that’s capitalism someone will.} & \makecell{This tweeter is pro-vaccine and says that he is pro-vaccine.}\\
\hline
\makecell{@USER Also Neil I love the way you throw about the word\\  vaccines, I know vaccines work.. I am pro vaccine.. My problem\\ is specifically the Covid19 jab.. I know ye like to throw everyone \\in the anti vaxx bucket.. $<sep>$ The message communicated\\ in this tweet is pro-vaccine. $<sep>$ This tweet criticises \\vaccine mandates. $<sep>$ This tweet supports or promotes\\ freedom of choice $<sep>$ This tweet supports or\\ promotes vaccines $<sep>$ This tweet supports or promotes\\ a more relaxed approach.} & \makecell{This tweeter says they are pro vaccines and that they are against the vaccine \\mandates}\\

\hline

\end{tabular}
\end{table*}

\subsection{Discussion on Available Subset}
Since a proportion of the tweets may not be accessible on Twitter temporarily or permanently due to tweet removal or user account deactivation, we create a subset of tweets that were still available around 9 months after the Twitter data collection was concluded. This accounts for around 73\% of the tweets. We compare the main statistics and results on this still-available subset, with the complete dataset; hereinafter referred to as $A$ and $C$, respectively. For the still-available subset, we follow the same train, validation, and test data split procedure as previously described for the complete set.

We report the results of RoBERTa-large on the classification tasks on the still-available subset and compare them with the results on the complete dataset in Table \ref{tab_t9}. RoBERTa-large is selected due to its overall performance regarding the comparisons on the complete dataset. 
Surprisingly, although the number of tweets has decreased in the still-available subset, the performance was increased on communicated message prediction task across all metrics. The increase in performance was observed on other models too. This improvement may be due to the decrease in the proportion of anti-vaccine tweets from 43.8\% to 38.6\% and the subsequent increase in the ratio of pro-vaccine tweets from 40.5\% to .45.4\%. This interesting change may be a sign of revision in the attitude of anti-vaxxers over time.  Although the results of RoBERTa-large were improved also on the misleading/inaccurate information detection task, this improvement was not observed in other models showing the capability of RoBERTa-large and the importance of model selection in this task. Also, the proportion of tweets with misleading or inaccurate information decreased from 38.9\% to 34.2\%. This may be related to an increase in public awareness over time.

The performance on subjects of criticism prediction and subjects of support/promote prediction tasks has decreased compared to the performance on the complete dataset. This decline was 2.7\% and 3.4\% in terms of average F1 score on these tasks, respectively. A similar trend was observed for the other models too. Regarding the ratio of positive class, while this number was similar for most of the labels, the highest increase was 1.4\% belonged to anti-vaxxers while the highest decline was 3.3\% held by vaccine safety. On subjects of support/promote prediction task, the maximum rise and decline in the positive class ratio were 4.3\% for vaccine and 3\% for freedom of choice, respectively. As for the tweet meaning generation task, we did not observe a noticeable difference between the results of the still-available subset and the complete dataset.

\begin{table*}
\caption{Comparison between the performance of RoBERTa-large on the complete dataset ($C$) and still-available subset ($A$) on the classification tasks.}
\label{tab_t9}
\centering
\begin{tabular}{l|cc|cc|cc|cc}
\hline
\multirow{2}{*}{Task} & \multicolumn{2}{c}{Accuracy} & \multicolumn{2}{c}{Precision} & \multicolumn{2}{c}{Recall} & \multicolumn{2}{c}{F1} \\

& T & A & T & A & T & A & T & A\\
\hline
Communicated Message Prediction & 69.6 & 71.7 & 61.9 & 65 & 61.8 & 63.6 & 61.8 & 64\\

Misleading Information Detection & 75.6 & 76.5 & 68 & 68.8 & 73.7 & 75.3 & 70.7 & 71.9\\

Subjects of criticism prediction & 89.2 & 89.9 & 51.6 & 48.3 & 41.2 & 39 & 44.7 & 42\\

Subjects of support/promote prediction & 88.8 & 88.5 & 38.6 & 33.9 & 29.5 & 26.6 & 32.8 & 29.4\\

\end{tabular}

\end{table*}

\section{Conclusion}
In this paper, we presented a novel Twitter COVID-19 vaccine misinformation dataset collected using Twitter API and manually annotated by a team of annotators with a background in communications and journalism. Each tweet of the dataset accompanies by an extensive set of human-provided annotations including vaccine hesitancy stance, indication of any misinformation in tweets, the entities criticized and supported in each tweet and the communicated message of each tweet. We also defined five baseline tasks on our dataset and reported the results of a set of recent NLP models for them. Furthermore, the provided dataset can be used to study the vaccine hesitancy and spread of vaccine misinformation problems.

\bibliographystyle{IEEEtran}
\bibliography{paper}

\end{document}